\documentclass[prd,aps,twocolumn,showpacs,superscriptaddress,amsmath,floatfix,amssymb,preprintnumbers]{revtex4}

\usepackage{amsmath, amstext,amssymb,amsfonts, graphicx,color,subfigure, multirow}
\usepackage{hyperref}

\newcommand{\ddd}{\displaystyle}

\begin{document}
\preprint{ITEP-LAT/2011-12}
\preprint{DESY 11-176}
\sloppy

\title{Fractal dimension of the topological charge density distribution\\ in SU(2) lattice gluodynamics}

\author{P.V. Buividovich}
\affiliation{JINR, Dubna, Moscow region, 141980 Russia}
\affiliation{ITEP, B. Cheremushkinskaya 25, Moscow, 117218 Russia}

\author{T. Kalaydzhyan}
\affiliation{DESY, Notkestrasse 85, 22607 Hamburg, Germany}
\affiliation{ITEP, B. Cheremushkinskaya 25, Moscow, 117218 Russia}

\author{M.I. Polikarpov}
\affiliation{ITEP, B. Cheremushkinskaya 25, Moscow, 117218 Russia}

\date{\today}
\begin{abstract}
We study the effect of cooling on the spatial distribution of the
topological charge density in quenched SU(2) lattice gauge theory with overlap fermions. 
We demonstrate that as the gauge field configurations are cooled, the Hausdorff dimension of regions where
the topological charge is localized gradually changes from $d=2 \div 3$
towards the total space dimension. Therefore, the cooling procedure
destroys some of the essential properties of the topological charge
distribution.
\end{abstract}

\maketitle

\section*{Introduction}

Topological charge density is an important characteristic of the
QCD vacuum, recently involved in phenomenological studies of many new hypothetical effects \cite{CME,phenomenology,phenomenology1,phenomenology2,phenomenology3,superfluidity}.
However, the spatial structure of the topological density distribution seems to be not well defined
since the relevant properties of the underlying vacuum structure depend on the measuring procedure \cite{Zakharov:2006te, Forcrand}.
The classical instanton approach \cite{'tHooft:1976fv} assumes that the nonperturbative physics is governed by the scale of $\Lambda_{QCD}$,
which means that the dimensionful quantities like volumes occupied by topological fermion modes should depend on $\Lambda_{QCD}$
 but not on the lattice spacing. On the contrary, the lattice measurements demonstrate that these volumes do depend on the spacing
(i.e. on the measurement resolution) and shrink to zero in the continuum limit \cite{Aubin:2004mp, Bernard, Gubarev:2005jm, Ilgenfritz:2007xu, Ilgenfritz:2007b}.

\begin{figure}[t]
\centering
\includegraphics{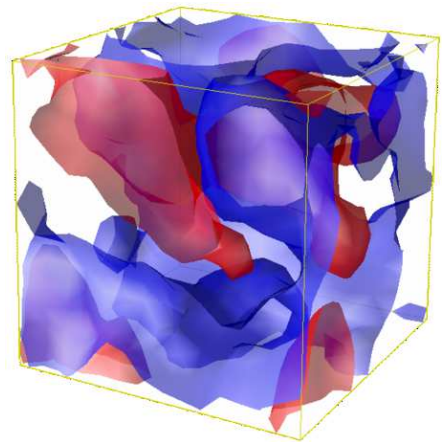}
\caption{Isosurfaces of the topological charge density\\ $q(x)=\pm 10^{-4}$ for a fixed time slice, corresponding to the $16^4$ lattice in Table~\ref{lattices}. Colors represent positive (red) and negative (blue) values, respectively. For the animation, see \cite{animation}.}
\label{distribution}
\end{figure}

It turns out that the continuum definition of the topological charge density
\begin{align}
q(x) = \ddd\frac{1}{32\pi^2}\epsilon^{\mu\nu\alpha\beta}\, \mathrm{Tr}\left(G_{\mu\nu}^a\, G_{\alpha\beta}^a\right)
\label{topcharge}
\end{align}
cannot be directly applied to the lattice gauge theory, since the
discretized version of (\ref{topcharge}) is no longer a full derivative. There are
two widely used methods to study the topology of gauge fields on the
lattice. First, one can apply a smearing procedure, which makes the
gauge fields smoother and thus closer to the classical fields. Second,
one can rely on the lattice version of the Atyah-Singer theorem and
define the total topological charge of a gauge field configuration
as the number of zero modes of the overlap Dirac operator \cite{Neuberger}
on this configuration. The corresponding local density of topological
charge can be defined, for example, as follows \cite{Horvath:2002yn, Luscher:1998pqa, Hasenfratz:1998ri}:
\begin{align}
q(x) = - \mathrm{Tr} \left[\gamma_5 \left( 1 - \frac{a}{2} D(x, x)\right) \right]\,,
\label{latttopcharge}
\end{align}
where $D(x, x)$ is the zero-mass Neuberger operator and the trace is taken over spinor and color indices.
Another attractive property of this definition is that it allows us to measure a \textit{local} imbalance 
in the number of left- and right-handed quarks (chirality), which is important for lattice studies of the 
\textit{local} CP-violation in strong interactions \cite{Kharzeev}. A typical result of the lattice 
simulation for this quantity (without cooling) is shown in Fig.~\ref{distribution}.

At the moment there are many investigations related to the spatial
structure of the topological charge distribution \cite{Ilgenfritz:2008ia, cooling, Bruckmann:2011ve, Ilgenfritz:2007xu, Ilgenfritz:2007b, Kovalenko:2005rz, Horvath}, which use both
of the alternative definitions. The measurements which rely on the cooling
procedure mostly suggest an instanton-like picture of the QCD vacuum
 \cite{Leinweber:1999cw}, while the definition (\ref{latttopcharge}) typically shows that the
topological charge is localized at low-dimensional objects
(defects) \cite{Ilgenfritz:2007xu, Ilgenfritz:2007b, Kovalenko:2005rz, Horvath} and has a very-long-range structure of the
distribution \cite{Horvath}. At the qualitative level it is known that both
definitions yield the topological charge densities which are strongly
correlated \cite{Ilgenfritz:2008ia, Zhang, Bruckmann:2006wf}. For an alternative filtering method based on adjoint fermions see Ref.~\cite{add}.

The aim of this paper is to fill the existing gap in the
literature and to demonstrate in what way the cooling procedure affects the
dimensionality of regions where the topological charge
density is localized. We use the definition (\ref{latttopcharge}) based on zero modes of
overlap Dirac operator and show that as the gauge field configurations
are cooled the dimension of these regions gradually tends to 4,
which is the total space dimension. The procedure makes the effective resolution of the measurement lower and thus provides a
 result close to the instanton picture. We verify our result using
several measures of the localization \cite{Aubin:2004mp, Gubarev:2005jm, Gattringer}.

\begin{center}
\begin{table}
 \begin{tabular}{|c | c | c | c | c|}
\hline
  $\beta$ & $a$ [$\mathrm{fm}$] & $L_s^3 \times L_t$ & $V$ [$\mathrm{fm}^4$] & \# conf \\ \hline \hline
3.200  & 0.117 & $12^3\times12$ & 3.93 & 50$\times$13 \\ \hline
3.295  & 0.100 & $14^3\times14$ & 3.90 & 50$\times$13 \\ \hline
3.332  & 0.094 & $15^3\times15$ & 3.89 & 50$\times$13 \\ \hline
3.365  & 0.088 & $16^3\times16$ & 3.88 & 50$\times$13 \\ \hline
3.425  & 0.078 & $18^3\times18$ & 3.87 & 50$\times$13 \\ \hline
\end{tabular}
\caption{Lattice parameters used in the calculation: couplings $\beta$, lattice spacings $a$, lattice sizes $L_s^3 \times L_t$, physical volume $V$, and number of gauge field configurations.}
\label{lattices}
\end{table}
\end{center}
\vspace{-1.2cm}
\section*{Technical details}
We work in the quenched $SU(2)$ lattice gauge theory with the tadpole-improved Wilson-Symanzik action \cite{Bornyakov:2008bg}. Lattices we used are listed in Table \ref{lattices}. We also implement the cooling procedure described in Ref.~\cite{cooling} with coefficient $c=0.5$ for the APE-smearing. For each lattice spacing we consider thirteen different stages of the cooling procedure: 0, 1, 2, 5 - 12, 20 and 50 iterations of the algorithm.
For valence quarks we use the Neuberger's overlap Dirac operator \cite{Neuberger}. Its eigenvalues and eigenfunctions are given by the following relation
\begin{align}
D \psi_\lambda = \lambda\, \psi_\lambda\,.
\end{align}

The quantities we measure in the present work are functions of two basic ingredients: the ``chiral condensate''
computed on a mode with eigenvalue $\lambda$,
\begin{align}
\rho_\lambda(x) = \psi_\lambda^{* \alpha}\, (x)\,\,\psi_{\lambda \alpha}(x)
\end{align}
and ``chirality'' computed on a mode with eigenvalue $\lambda$ [in agreement with the definition (\ref{latttopcharge})],
\begin{align}
\rho_\lambda^5(x) = \left(1 - \frac{\lambda}{2}\right)\psi_\lambda^{* \alpha}\, (x)\gamma^5_{\alpha\beta}\, \psi_\lambda^\beta(x)\,.
\end{align}
Here we sum over spinor and (omitted) color indices.
The total values of both chiral condensate and chirality are given by an infinite sum over all eigenvalues. Lattice studies \cite{DeGrand:2000gq, DeGrand:2001tm} suggest that the long-distance properties of QCD can be treated with a finite cutoff of the fermionic spectrum. We hereby restrict our consideration to the IR part of the Dirac spectrum consisting of zero modes ($\lambda=0$)
 and few low-lying modes ($\lambda\neq0$).

Inverse participation ratio (IPR) for an arbitrary normalized distribution $\alpha(x)$ is usually defined in the following way
\begin{align}
\mathrm{IPR} = \left\{ N \sum\limits_{x} \alpha^2(x) \left|\, \sum\limits_x \alpha(x) = 1 \right. \right\}\,,\label{IPR}
\end{align}
where $N$ is the total number of lattice sites $x$. From this definition one can clearly see that $\mathrm{IPR} = N$ if $\alpha(x)$ is localized on a single site and $\mathrm{IPR} = 1$ if $\alpha(x) = const$, i.e. the distribution is unlocalized. In general $\mathrm{IPR}$ is equal to the inverse fraction of sites occupied by the support of $\alpha(x)$.
Since this fraction of sites can be thought of as a number of
four-dimensional lattice hypercubes covering the support, the Hausdorff dimension $d$ of these regions can be extracted from the asymptotic behavior of IPR at small
lattice spacings $a$
\begin{align}
 \mathrm{IPR}(a) = \ddd\frac{c}{a^{d}}\label{fit}\,,
\end{align}
where $c$ is a constant. It is also useful to mention, that in physical units $\mathrm{IPR}^{-1}$ is equal to the part of the total volume occupied by the distribution.

In the following sections we will modify the standard definition (\ref{IPR}) to adapt it to our particular cases (i.e. unnormalized or non-normalizable distributions, etc.). The final result will show an equivalence of the chosen definitions.

\subsection*{Ordinary IPR for zero modes.}
\begin{figure*}[!ht]
\vspace{-1cm}
\subfigure{\includegraphics[width=5.6cm, angle=-90]{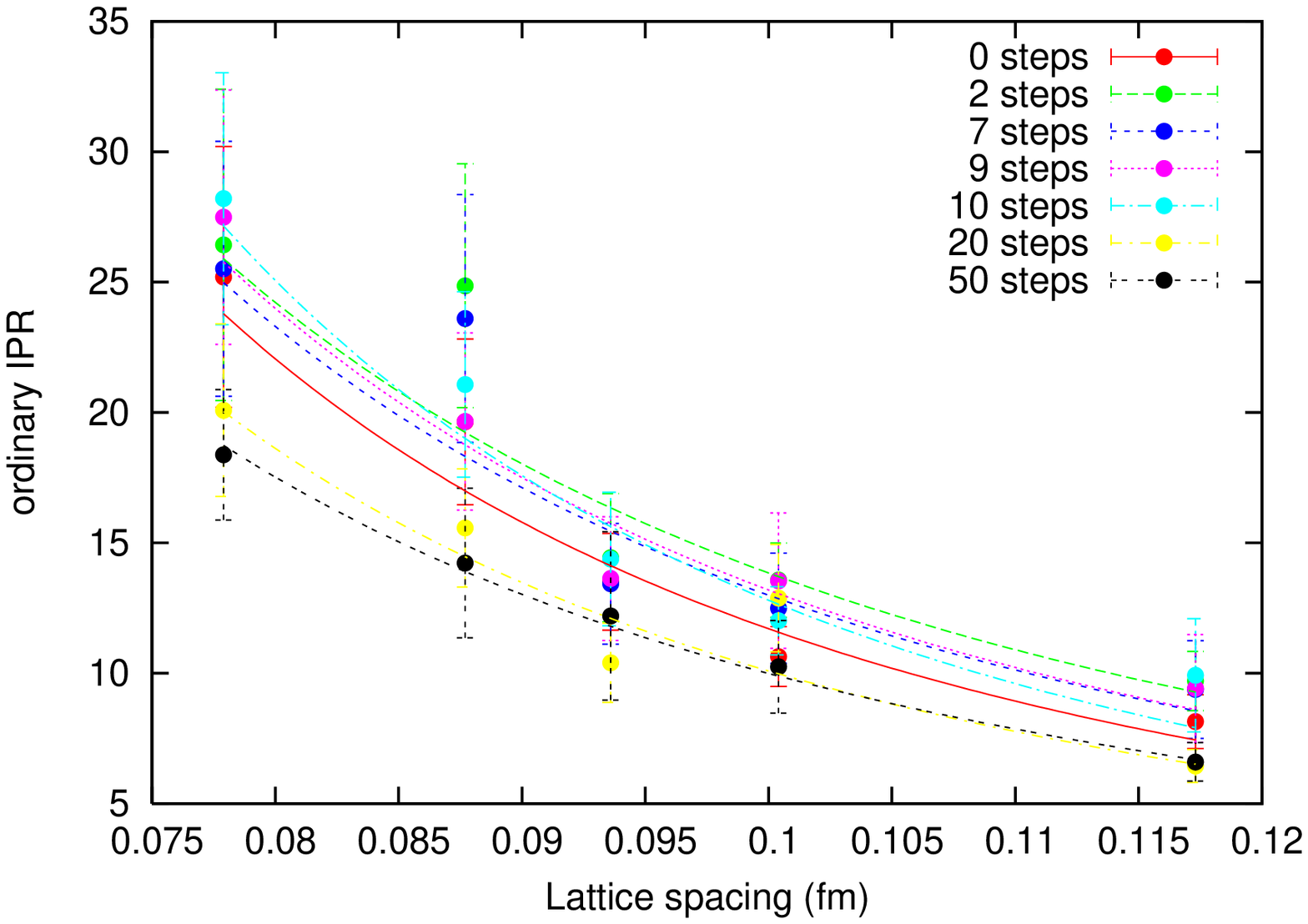}}
\subfigure{\includegraphics[width=5.6cm, angle=-90]{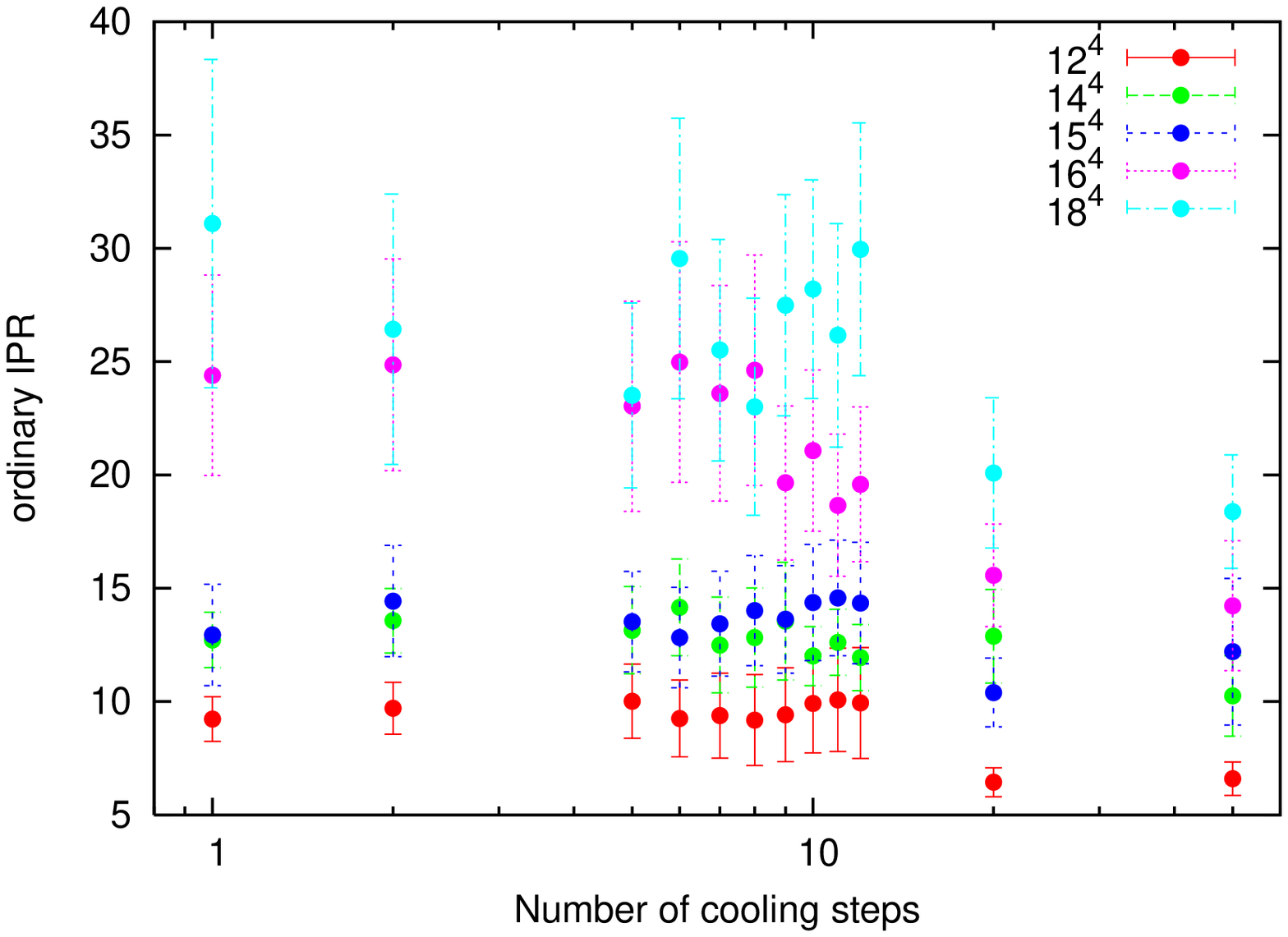}}
\vspace{-0.3cm}
\caption{Ordinary IPR for zero modes (\ref{DEF1}).}\label{gubarev_ipr}
\subfigure{\includegraphics[width=5.6cm, angle=-90]{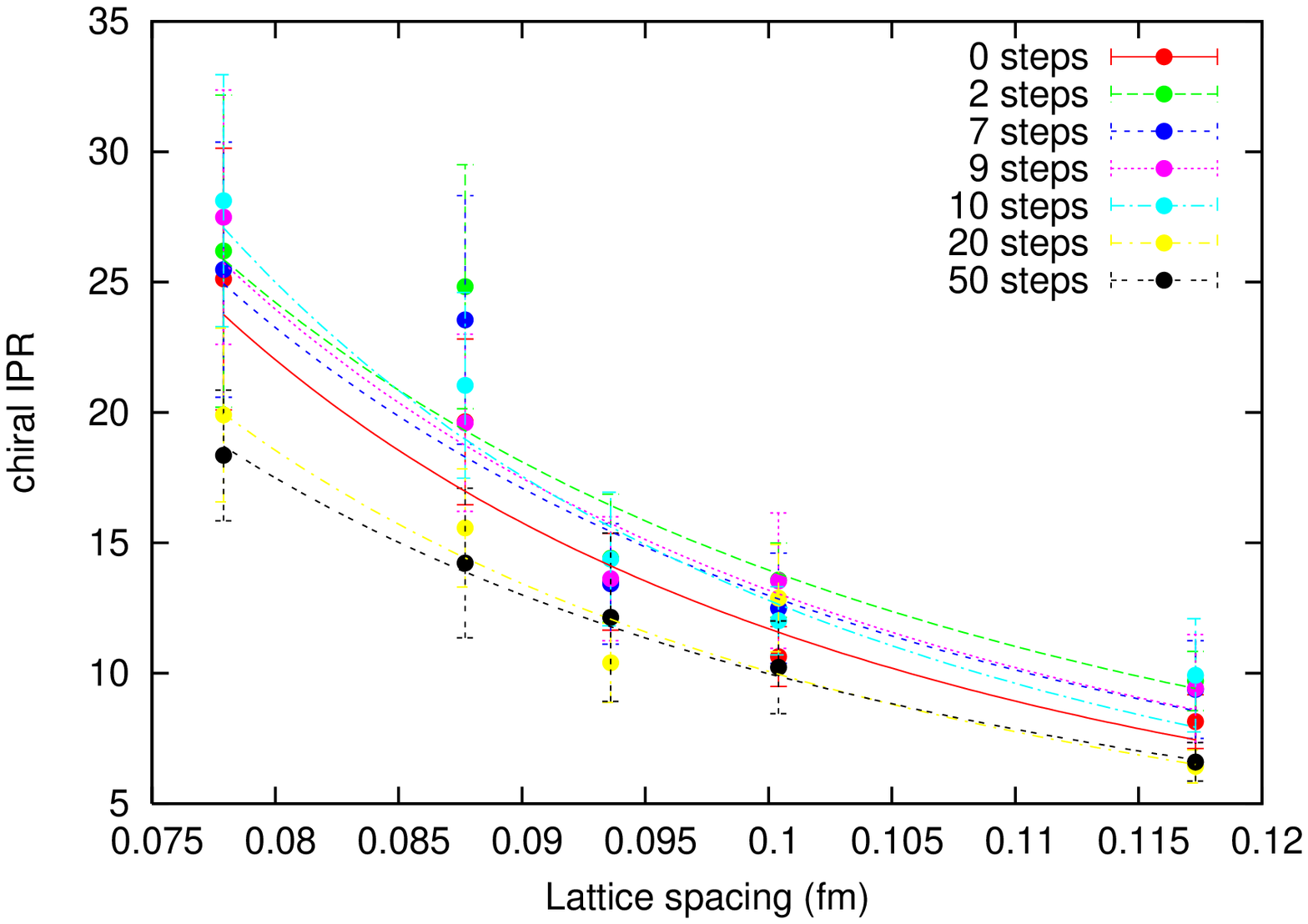}}
\centering
\subfigure{\includegraphics[width=5.6cm, angle=-90]{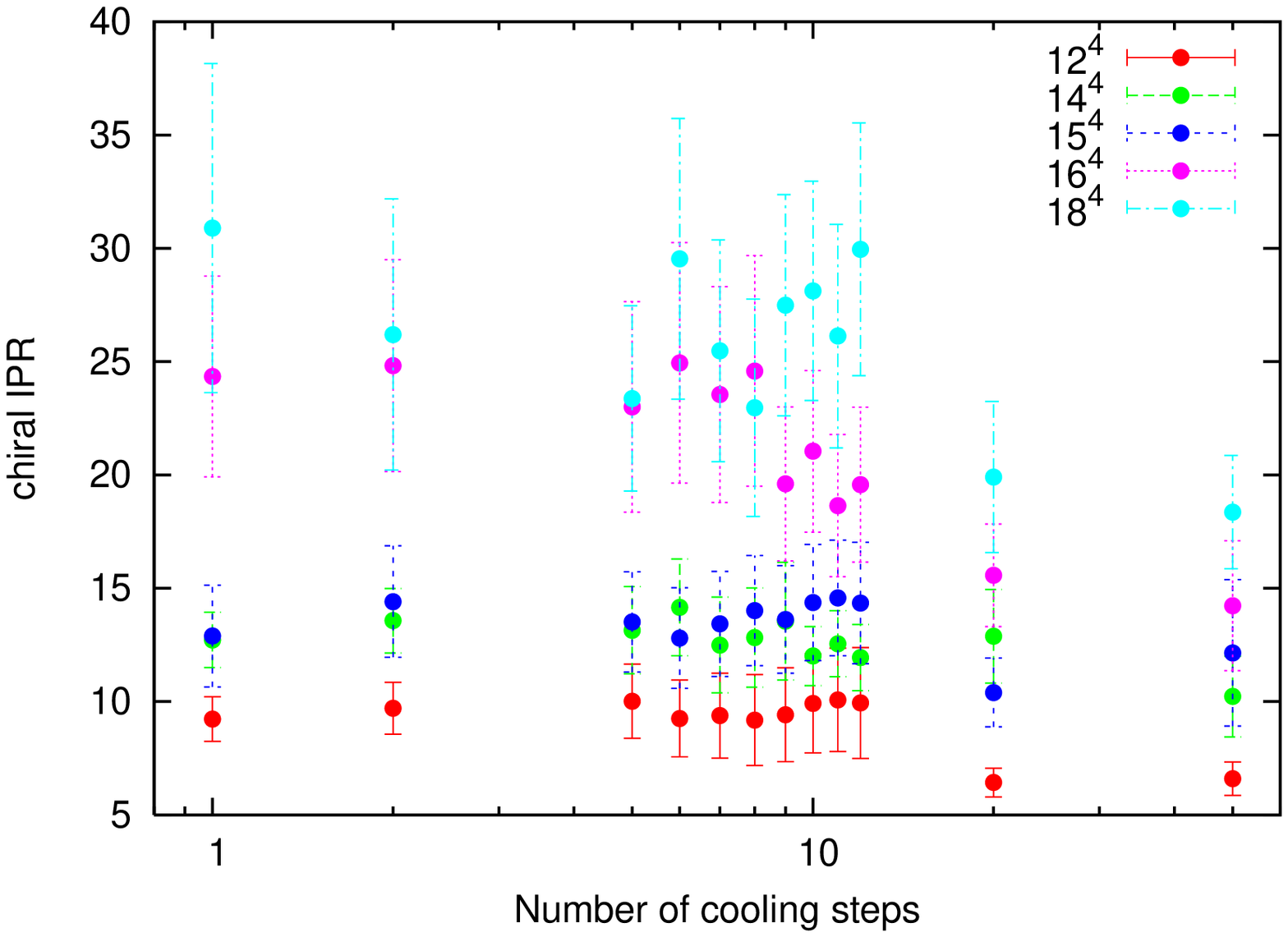}}
\vspace{-0.3cm}
\caption{Chiral IPR for zero modes. First definition, Eq.~(\ref{DEF2}).}\label{gattringer_ipr1}
\subfigure{\includegraphics[width=5.6cm, angle=-90]{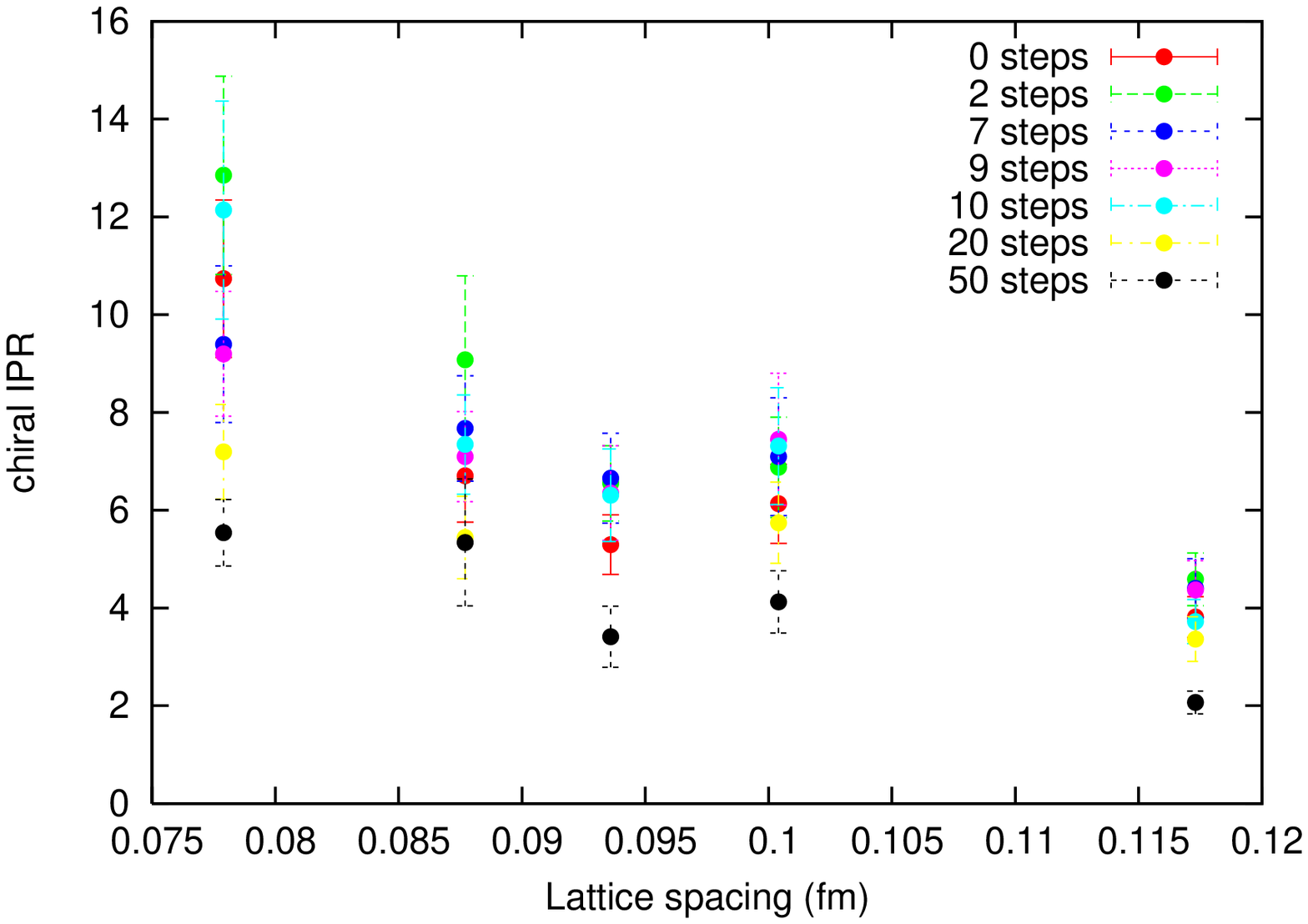}}
\centering
\subfigure{\includegraphics[width=5.6cm, angle=-90]{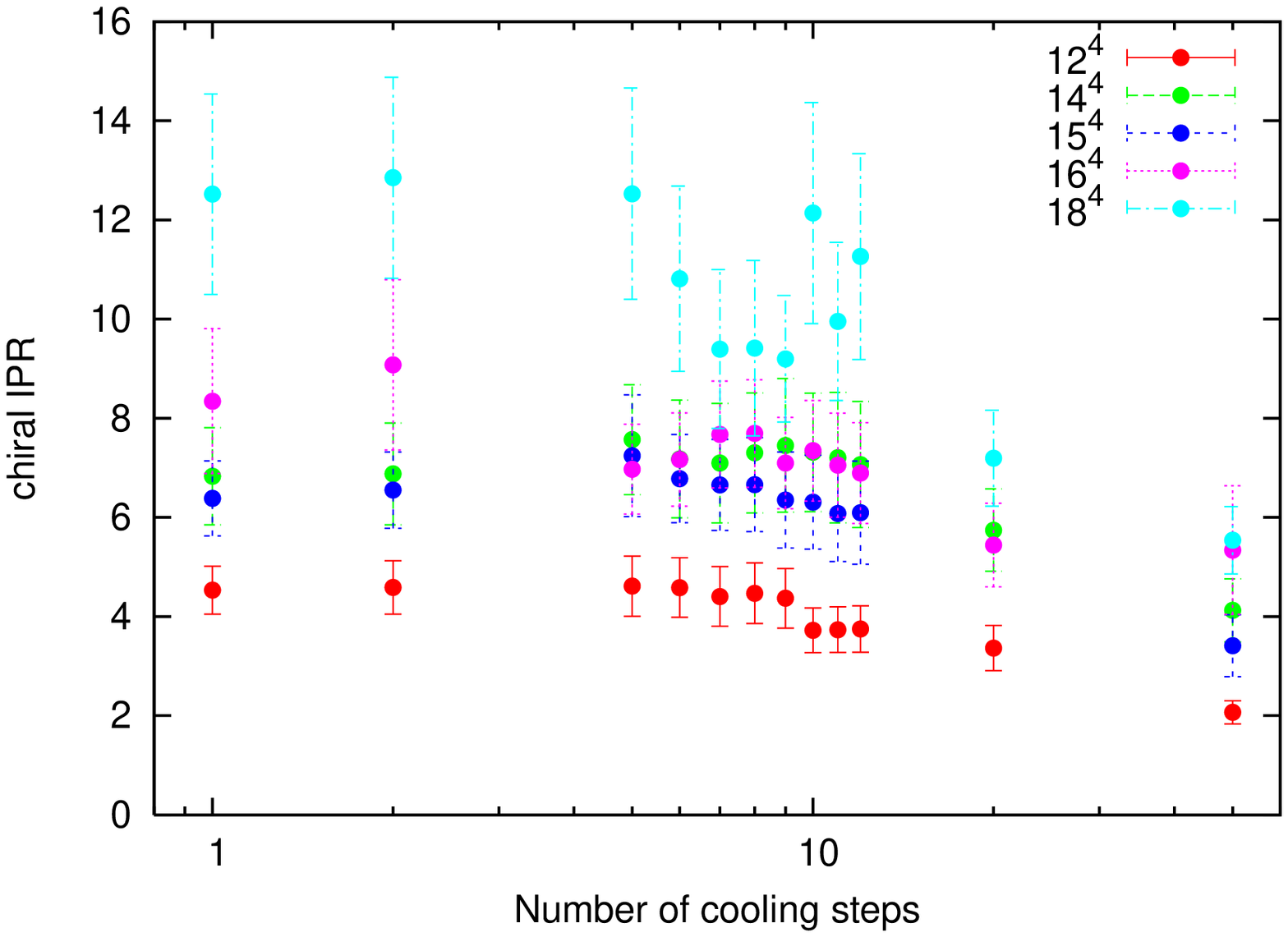}}
\vspace{-0.3cm}
\caption{Chiral IPR for the lowest nonzero modes. First definition, Eq.~(\ref{DEF3}).}\label{gattringer_ipr2}
\subfigure{\includegraphics[width=5.6cm, angle=-90]{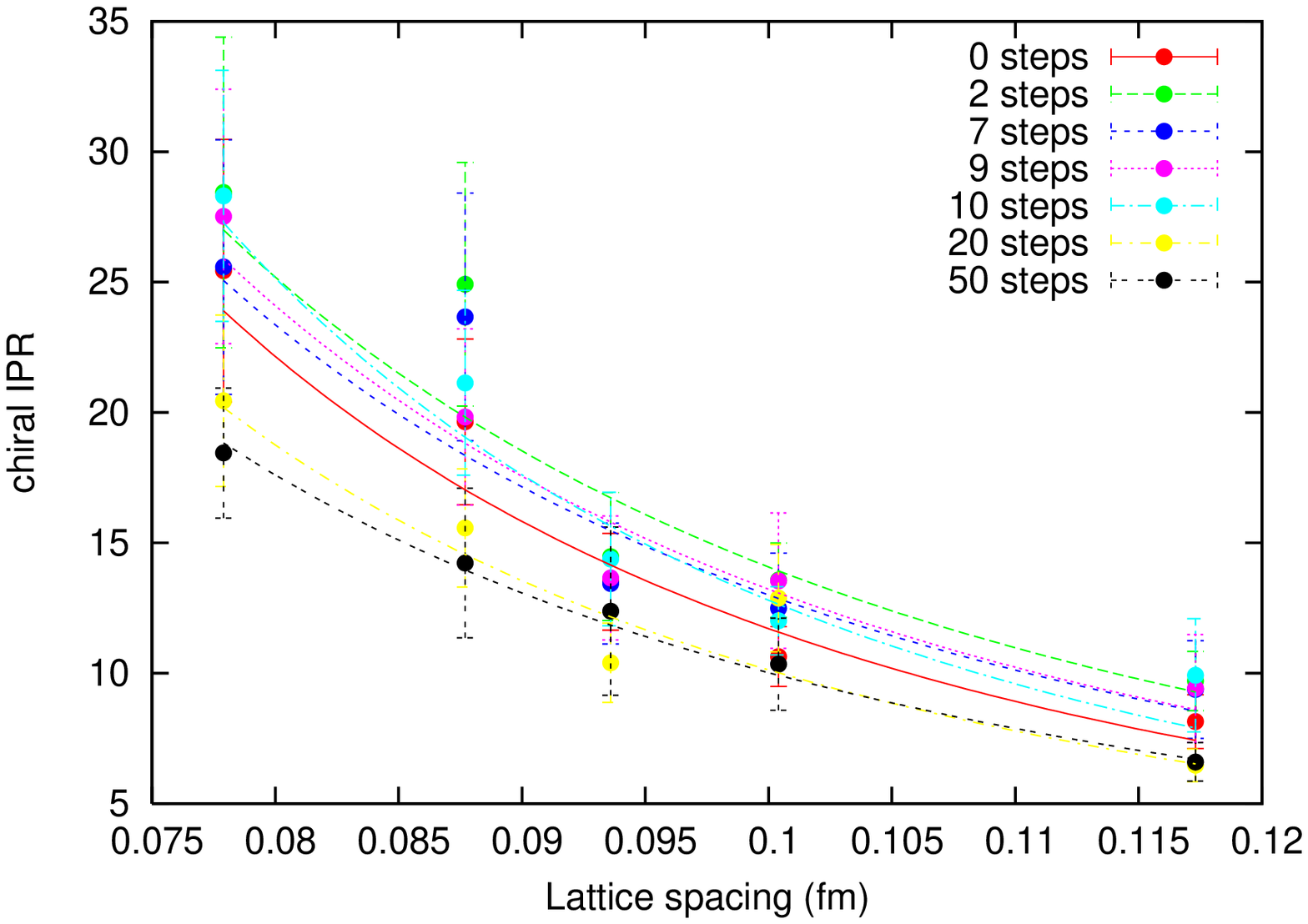}}
\centering
\subfigure{\includegraphics[width=5.6cm, angle=-90]{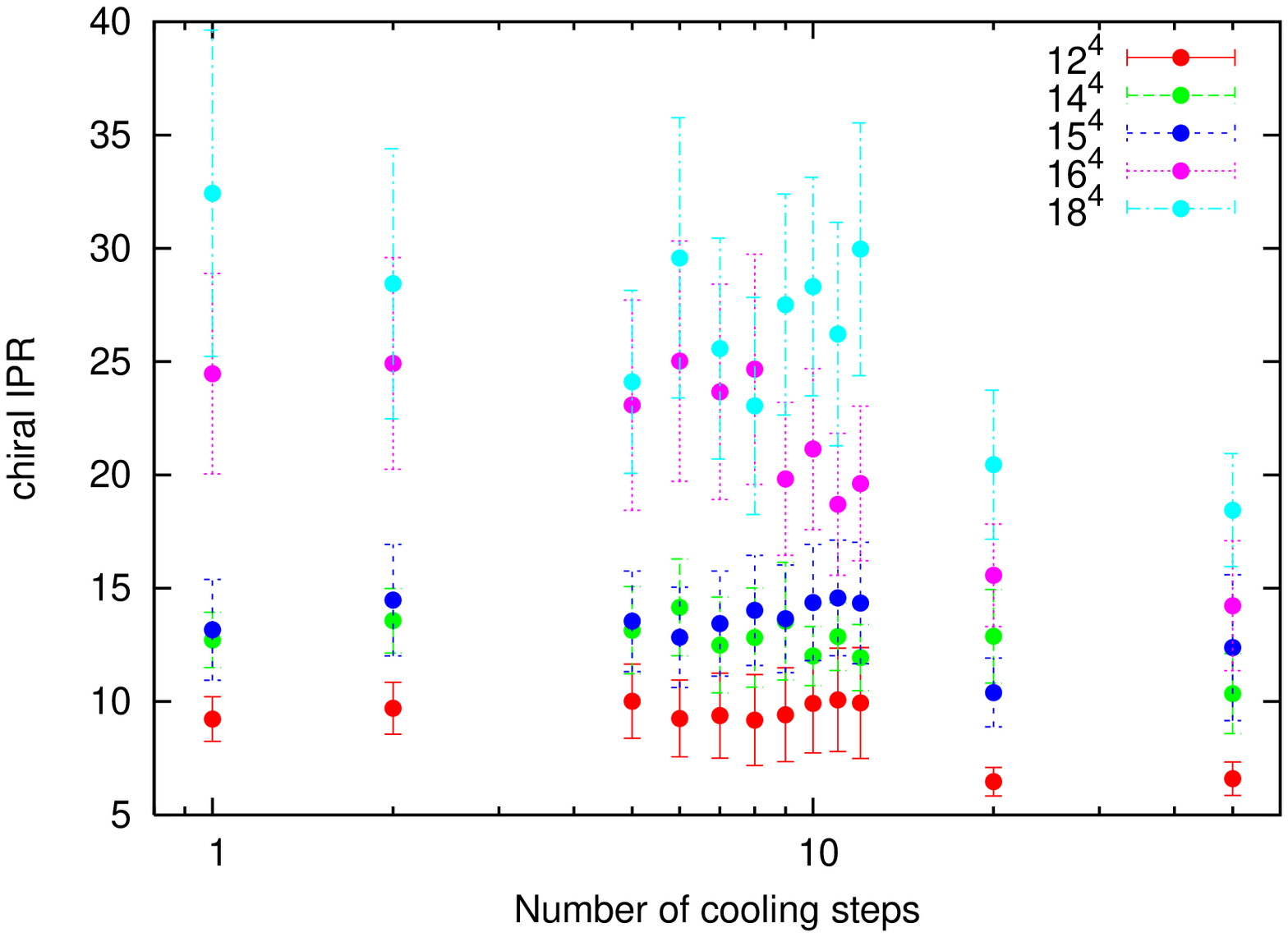}}
\vspace{-0.3cm}
\caption{Chiral IPR for zero modes. Second definition, Eq.~(\ref{DEF4}).}\label{aubin_ipr}
\end{figure*}

In this section we compute the inverse participation ratio for the fermionic zero modes according to the one defined in Ref.~\cite{Gubarev:2005jm}:
\begin{align}
\mathrm{IPR_0} = N \ddd\left[\frac{\ddd\sum\limits_x  \left(\rho_0(x)\right)^2}{\left(\ddd\sum\limits_x \, \rho_0(x)\right)^2}\right]_{\lambda=0}\,,\label{DEF1}
\end{align}
where the brackets $[...]_{\lambda=0}$ denote an averaging over all zero modes and further averaging over all gauge field configurations. Results are presented in Fig.~\ref{gubarev_ipr}. 

The left-hand figure shows how the localization depends on the lattice spacing $a$ - the finer the lattice, the larger the IPR.
This fits very well to the idea of vanishing total volume occupied by fermionic zero modes in the continuum limit $a \rightarrow 0$ (see Ref.~\cite{Zakharov:2006te} for a review). Using the fit (\ref{fit})
we recover the fractal (Hausdorff) dimension $d$ of the volume. Results for the fits with fixed numbers of cooling steps are presented in the Table \ref{ipr_table}. Here, to minimize errors, we also prepared an alternative sample consisting only of those configurations which do not lose all the fermion zero modes during the cooling. We picked then the values with better (and also sufficient) statistical significance.
\begin{center}
\begin{table}
\vspace{.5cm}
\begin{tabular}{|c | c | c | c|}
\hline
Number of & Fractal& Standard & \multirow{2}{*}{P-value}\\
cooling steps &$\quad$ dimension $\quad$& error & \hspace{2cm} \\ \hline \hline
 0 & 2.84 $\pm$ 0.44 & 15\% & 0.008\\ \hline
 1 & 2.66 $\pm$ 0.66 & 25\% & 0.027\\ \hline
 2 & 2.49 $\pm$ 0.46 & 18\% & 0.013\\ \hline
 5 & 2.17 $\pm$ 0.49 & 23\% & 0.021\\ \hline
 6 & 2.75 $\pm$ 0.66 & 24\% &  0.025\\ \hline
 7 & 3.17 $\pm$ 0.51 & 16\% &  0.009\\ \hline
 9 & 3.71 $\pm$ 0.34 & 9\% & 0.001\\ \hline
 12 & 3.88 $\pm$ 0.23 & 6\% &  $4\cdot10^{-4}$\\ \hline
\end{tabular}
\caption{Fractal dimension of the fermionic zero modes and, equivalently, of the topological charge distribution.}\label{ipr_table}
\end{table}
\end{center}

\subsection*{Chiral IPR for low-lying modes. First definition.}
In this section we modify the IPR to measure localization properties of the topological charge distribution. The average chirality $\left[ \ddd\sum\limits_x \, \rho_\lambda^5(x)\right]_\lambda$ is zero, therefore we have to use either the absolute value $|\rho_\lambda^5(x)|$ or the square $\left[\rho_\lambda^5(x)\right]^2$. Here we stick to the definition from \cite{Gattringer}, which in our terms has the following form
\begin{align}
\mathrm{IPR^5_0} = N \left[\ddd\frac{\ddd\sum\limits_x  \left(\rho_0^5(x)\right)^2}{\left(\ddd\sum\limits_x  \rho_0(x)\right)^2}\right]_{\lambda=0}\,.\label{DEF2}
\end{align}

Results are presented in Fig.~\ref{gattringer_ipr1}. From the plots we conclude that the topological charge distribution behaves similar to the zero modes, tending to occupy a vanishing volume in the continuum limit. We can also compute the chiral IPR for small but nonzero eigenvalues (in our case we pick first 7 eigenvalues, $\lambda \lesssim 200 \, \mathrm{MeV}$),
\begin{align}
\mathrm{IPR^5_{\lambda \neq 0}} = N \left[\ddd\frac{\ddd\sum\limits_x  \left(\rho_{\lambda}^5(x)\right)^2}{\left(\ddd\sum\limits_x \, \rho_{\lambda}(x)\right)^2}\right]_{\lambda \neq 0}\,.\label{DEF3}
\end{align}
Chiral IPR for these modes is small (Fig.~\ref{gattringer_ipr2}) and thus the topological charge distribution at this part of the spectrum is delocalized.

\subsection*{Chiral IPR for zero modes. Second definition.}
Finally we consider a second definition of the chiral IPR according to \cite{Aubin:2004mp}:
\begin{align}
\mathrm{IPR^5_0} = N \left[\ddd\frac{\ddd\sum\limits_x  \left|\rho_0^5(x)\right|^2}{\left(\ddd\sum\limits_x \, |\rho^5_0(x)|\right)^2}\right]_{\lambda=0}\,,\label{DEF4}
\end{align}
where, as before, $\rho_0^5(x)$ denotes the chirality on a zero mode (\ref{latttopcharge}). Results are presented in Fig.~\ref{aubin_ipr}. As can be seen from Figs.~\ref{gubarev_ipr}, \ref{gattringer_ipr1}, and \ref{aubin_ipr} the IPR for the zero modes and for the topological charge density on these modes are the same up 
to negligible deviations. Results of the fitting procedure coincide for these three cases and are shown in Table~\ref{ipr_table}.
The coincidence is not accidental, because for the zero modes $[D, \gamma^5] = 0$ and $ \gamma^5 |\psi_0\rangle = \pm|\psi_0\rangle $. This means that on a given mode $\rho_0(x)$ and $\rho^5_0(x)$ are equal to each other up to a sign.

\begin{figure}[t]
\centering
\includegraphics[width=6cm, angle=-90]{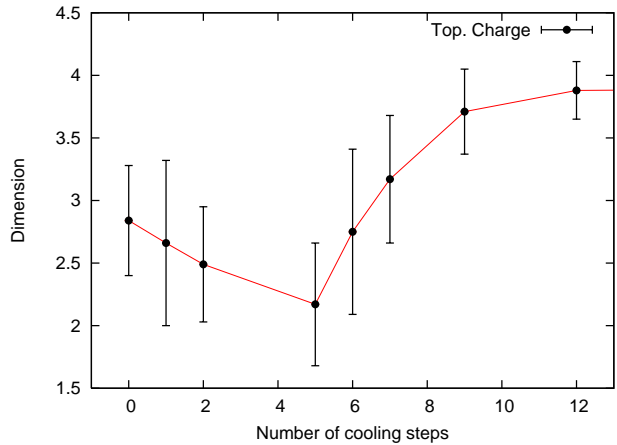}
\caption{Fractal dimensions at various cooling stages. The solid line is shown to guide the eye.}\label{dimensions_plot}
\end{figure}

\subsection*{Fractal dimension. Results and conclusions.}
\begin{figure}[t]
\centering
\includegraphics[width=6cm, angle=-90]{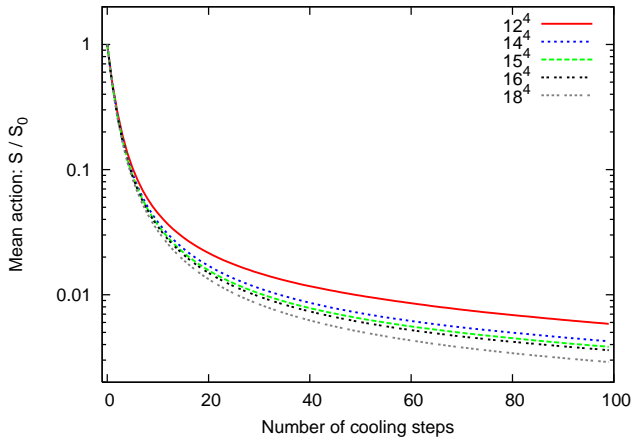}
\caption{Mean action at various cooling stages.}\label{action}
\end{figure}

To conclude, we demonstrate that the topological charge is localized on
low-dimensional fractal structures, whose fractal (Hausdorff) dimension depends on the number of cooling steps.
The obtained dimension is about $d = 2 \div 3$ for a few ($n<6$) steps of the cooling, while it grows to $d = 4$ with further iterations (see Fig.~\ref{dimensions_plot}). For a long cooling ($n\gtrsim20$) the result becomes insignificant, because the procedure leads to a delocalization of the distributions as can be seen in Figs.~\ref{gubarev_ipr}--\ref{aubin_ipr} (otherwise IPR remains consistent with a constant within error bars). We suppose that it can be caused by the annihilation of the instanton/anti-instanton pairs. Indeed, comparing the mean action evolution (Fig.~\ref{action}) with the one from e.g. Ref.~\cite{Polikarpov:1987yr} we see that the annihilation phase in our case could start already from $n \sim 20$. In Ref.~\cite{cooling}, where the same cooling algorithm is used, the annihilation takes place even at a smaller number of steps. 

The main conclusions of our paper are the following:
\begin{itemize}
 \item[(1)] Fermionic zero modes and chirality are localized on structures with fractal dimension $d = 2 \div 3$, which is
 an argument in favor of the vortex/domain-wall nature of the localization \cite{Zakharov:2004jh, Kovalenko:2004xm}.
 \item[(2)] A long sequence of iterations of the cooling procedure provides a
 result close to the instanton picture, i.e. destroys the low-dimensional structure of the QCD vacuum.
\end{itemize}

Finally, let us briefly mention a possible phenomenological consequence 
of our study. One of the most promising effects appearing due to the
nontrivial topology of the QCD vacuum is the so-called ``chiral magnetic effect'' (CME) \cite{CME},
which states the generation of an electric current in parallel
to an external magnetic field. Topological charge density in this case can
be understood as an imbalance in the number of left- and right-handed
light quarks induced by a nontrivial gluonic background. This
effect is expected to explain charge asymmetries observed at RHIC \cite{Kharzeev, Selyuzhenkov}.
Some evidences of the CME on the lattice as well as numerical estimates for the values of the local topological charge were also obtained in Refs.~\cite{Buividovich:2009wi,conductivity,Abramczyk:2009gb,Braguta:2010ej,Yamamoto:2011gk}.
At the current level of analytic studies CME is considered as an effect on the background of
spatially homogeneous axial fields \cite{Kharzeev:2010ym}, while the lattice simulations predict an irregular
structure of the would-be axial field (see Fig.~\ref{distribution}).
This spatial inhomogeneity can be treated within a chiral superfluid model \cite{superfluidity},
where the chirality is carried by an effective axion-like field. 
Knowledge of the nature of the topological charge localization can help us to translate lattice
Euclidean properties of the chirality to the language of an effective Minkowski field theory \cite{Chernodub:2009rt}.

\vspace{0.4cm}
\section*{Acknowledgements}
This work was supported by the grant for Leading Scientific Schools NSh-6260.2010.2 and RFBR 11-02-01227-a, Federal Special-Purpose Programme ``Cadres'' of the Russian Ministry of Science and Education. P.~Buividovich was supported by the postdoctoral fellowship of the
FAIR-Russia Research Center. The numerical calculations were performed at the GSI batch farm. Authors thank V.I.~Zakharov, I.~Horv\'{a}th, and P.~de~Forcrand for useful discussions.

\end{document}